\documentclass[10,iop,apjl]{emulateapj}
\usepackage{amssymb}
\usepackage{array}

\newcommand{\DxDy}

\begin{document}
\title{On the chemical mixing induced by internal gravity waves (IGW)}
\author{T.M. Rogers}
\affil{School of Mathematics, Statistics and Physics Newcastle University,
  UK}
\affil{Planetary Science Institute, Tucson, AZ 85721, USA}
\author{J.N.McElwaine}
\affil{Department of Earth Sciences, Durham University, UK}
\affil{Planetary Science Institute, Tucson, AZ 85721, USA}

\begin{abstract}
Detailed modeling of stellar evolution requires a better understanding
of the (magneto-)hydrodynamic processes which mix chemical elements and
transport angular momentum.  Understanding these processes is
crucial if we are to accurately interpret observations of chemical abundance
anomalies, surface rotation measurements and asteroseismic data.  Here, we use two-dimensional hydrodynamic
simulations of the generation and propagation of internal gravity
waves (IGW) in an intermediate mass star to measure the chemical
mixing induced by these waves.  We show that such mixing can generally be treated as a diffusive
process.  We then show that the local diffusion coefficient does not
depend on the local fluid velocity, but rather on the wave amplitude.  We
then use these findings to provide a simple
parametrization for this diffusion which can be incorporated into
stellar evolution codes and tested against observations.  
\end{abstract}

\section{Introduction}
Accounting for hydrodynamic processes in stellar interiors over
stellar evolution times has been the biggest source of
  uncertainty when comparing theoretical results with observations.  While Mixing Length Theory (MLT)
has proven extremely useful for characterizing mixing within
convection zones  \citep{bv58,kippenhahn}, there remain many uncertainties dealing with this mixing at
convective-radiative interfaces \citep{renzini87,zahn91} and within radiative
regions \citep{pin97,heger00}.  Nearly all stars host radiative regions so it is critical we develop
methods for accurately parametrizing chemical mixing (and angular momentum
transport) in these regions.  

Numerous theoretical models have been proposed for incorporating
mixing by (magneto-)hydrodynamical processes in stellar radiative
zones into stellar evolution models.  A myriad of hydrodynamic
instabilities \citep{heger00}, rotationally induced mixing
(Eddington-Sweet circulation, \citet{eddington25,Vogt25,sweet50}) and
magnetically induced mixing (Taylor-Spruit dynamo
\citet{spruit02}) have been included in modern stellar evolution
codes.  Still, many questions remain about mixing within stellar
radiative interiors.  For example, it is typically assumed that
rotationally induced mixing is dominant in massive stars, yet the 
  observed lack of correlation between Nitrogen abundance and rotation
  rate for some stars indicates additional mixing is needed \citep{brott11} and observations
in multivariate parameter space suggest pulsational mixing is dominant
for slow to moderately rotating OB stars \citep{aerts14}. Similarly, differential
rotation at late stages of stellar evolution is lower than expected
even when all of these mechanisms are considered \citep{eggenberger17}.

In general, these multi-dimensional hydrodynamic effects are
parametrized as a diffusion coefficient within stellar evolution
codes. Each physical process has an instability criteria based on
local properties (e.g. shear).  Once instability
is confirmed a diffusion coefficient is constructed from the
lengthscale and growth rate of the instability.  This diffusion
coefficient is then included locally (in space and time) in the
stellar evolution calculation.   While this procedure is rather
rudimentary it is clear from observations that
additional mixing within stellar radiative regions is required.  It is only recently, and in limited circumstances, that such
prescriptions are being tested against hydrodynamic calculations which self consistently
calculate the development of the instability and the subsequent mixing
induced \citep{edelmann17}.   

Internal gravity waves (IGW) are known to propagate and
dissipate in radiative regions which could lead to chemical mixing and angular
momentum transport.  However,  the parametrization of IGW in
one-dimensional (1D) stellar evolution codes is complex.  The
transport of angular momentum by these waves can not be treated as a diffusive process, indeed IGW
have an \textit{anti-diffusive} nature.  That is, they drive, rather
than dissipate, shear flows \citep{buhler09}.  For this reason, IGW transport has
generally not been treated in 1D stellar evolution codes (except in
the Geneva code \citep{talon05,charbonnel05}, in which their treatment
is complex). Yet, while it is clear that angular momentum transport by IGW can not
be parametrized with a diffusion coefficient \citep{rg15}, it is unclear whether
the \textit{chemical} mixing induced by waves could be treated
diffusively as previously suggested \citep{pressryb81,gls91}. 
The purpose of this letter is to first determine whether wave mixing can
be treated diffusively and, if so, determine how efficient that mixing
is and whether it could be reasonably parametrized in 1D stellar
evolution models.  
\section{Numerics}
\subsection{Hydrodynamic Simulations}
\begin{figure*}
\centering
\includegraphics[width=0.95\textwidth]{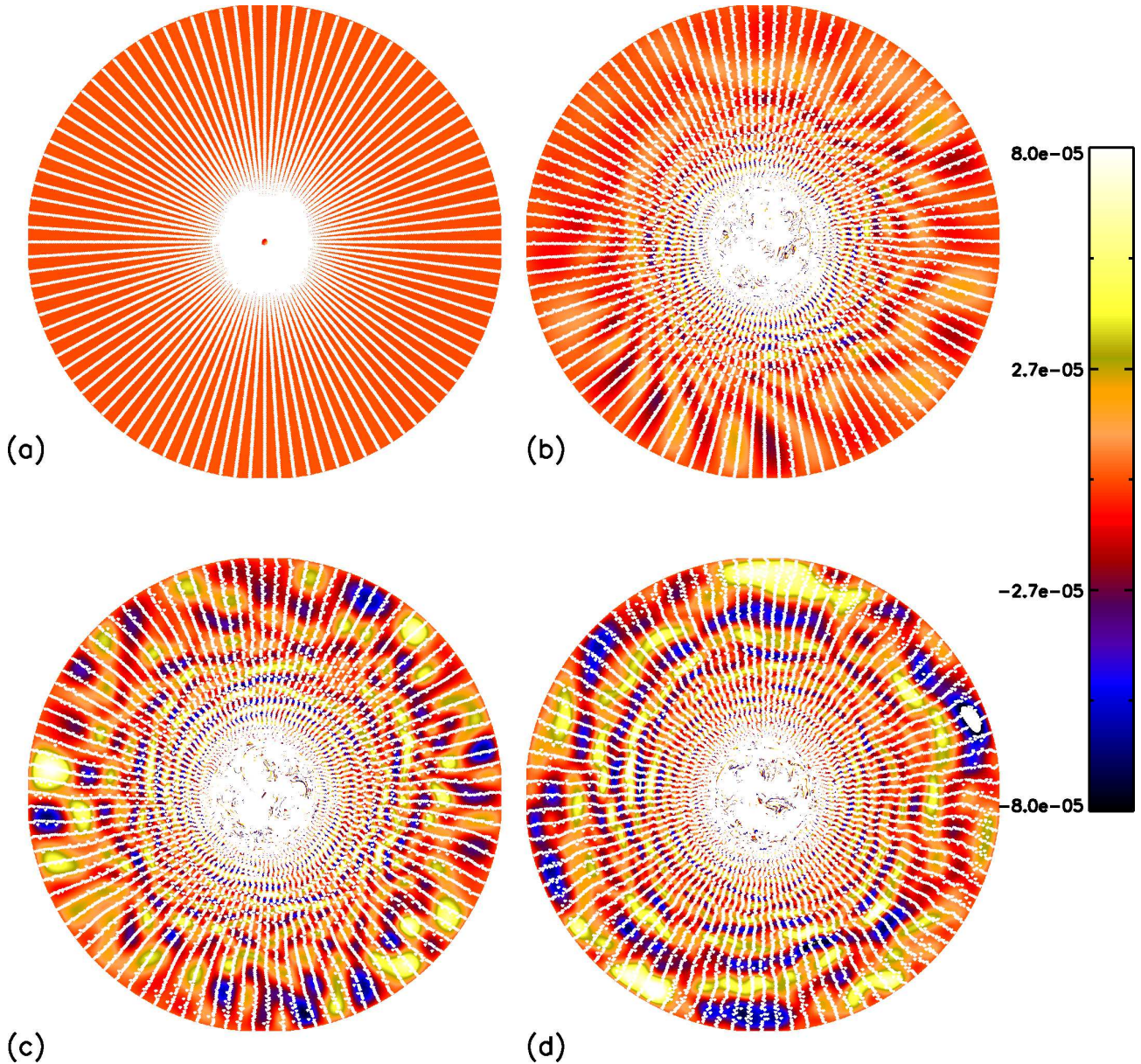}
\caption{Time snapshots of vorticity (color, units are rad/s) with
  particle positions, indicated with white dots overplotted.  Particle
  positions in (a) represent a subset of the initialized particles.
  One can clearly see movement of the particles, in response to wave
  motion, in the radiation zone in (b-d) and particles within the
  convection zone are fully mixed (difficult to discern as they cover
  the region).}
\label{snaps}
\end{figure*}

In order to measure mixing by IGWs in stellar interiors, we solve the
Navier-Stokes equations in the anelastic approximation
\citep{go69,rg05b}.  The equations are solved in two dimensions (2D)
representing an equatorial slice of the star.  Our reference state
model is that of a 3M$_{\odot}$ star with a core hydrogen content of
0.35, calculated using Modules for Experiments in Stellar Astrophysics
(MESA, \cite{paxton10,paxton15}).  We solve the equations from
0.03R$_{\ast}$ to 0.70R$_{\ast}$, the initial rotation rate is uniform
and equal to 10$^{-6}$ rad/s.  The simulation is run a total of
4$\times$10$^7$s, or approximately 40 convective turnover times. The
details of the equations and numerical methods used can be found in
\cite{rg13}. Time snapshots of vorticity are shown in Fig.~\ref{snaps}. Ideally, we would solve the equations in 3D as in
\citet{alvan15}, but few simulations such as these have been done for
massive stars \citep{browning04,augustson16} and none which include
extended radiative zones.  We discuss the role this reduced
dimensionality might play in the Discussion.

Like all hydrodynamic simulations our viscous and thermal diffusion
coefficients are larger than is typical in real stars. Therefore, our
waves are damped more than they would be in actual stellar
interiors. Unlike the simulations presented in \cite{rg12,rg13}, these
models are not ``over-forced'' so their root-mean-squared velocities
in the convection zone are similar to that expected from mixing length
theory (see Fig.~\ref{velvsdiff}).

\subsection{Tracer Particles and Diffusion}
To determine whether mixing by IGW behaves like diffusion we introduce
tracer particles within the simulation.  We use a tracer particles
instead of solving a compositional advection-diffusion equation
because such a method would require an explicit diffusion that would
dominate any mixing coefficient.  For simplicity we regard time as continuous and only
measure radial diffusion.  At some time of interest we
introduce $N$ particles, distributed uniformly in space, and track
them for a time $T$ to produce $N$ particle trajectories (particle
positions in time, $R_i(t)$). We then consider all the sub-trajectories of duration $\tau$ and use a
cubic spline function $w$ to interpolate between the start of each
sub-trajectory ($R_i(t)$) and the grid position. The length of a
sub-trajectory (displacement of a particle) is then
$R_i(t+\tau)-R_i(t)$.  We then calculate the following
profiles:
\begin{eqnarray}
  n(r,\tau)  &=& \int_0^{T-\tau} \sum_{i=1}^N w\left(R_i(t)-r\right)                      \,dt\\
  P (r,\tau) &=& \int_0^{T-\tau} \sum_{i=1}^N w\left(R_i(t)-r\right)[R_i(t+\tau)-R_i(t)]   \,dt\nonumber\\
  Q (r,\tau) &=& \int_0^{T-\tau} \sum_{i=1}^N w\left(R_i(t)-r\right)[R_i(t+\tau)-R_i(t)]^2 \,dt\nonumber
\end{eqnarray}
Here $n(r,\tau)$ is the number of sub-trajectories starting at $r$ of
duration $\tau$. $P\left(r,\tau\right)$ is the sum of lengths of
these sub-trajectories and $Q\left(r,\tau\right)$ is the sum of the
lengths squared of these sub-trajectories. If there is a mean velocity
field $u(r)$ then
\begin{equation}
u(r,\tau)=\frac{P(r,\tau)}{n(r,\tau)\tau}
\end{equation}
Lower values of $\tau$ can result from many time differences, while
larger values of $\tau$ can only result for long timeline
data.\footnote{As an example if the time step is 1 and $T$=100 a time
  difference of $\tau=10$ can result from 10--0, 11--1, 12-2 and so
  on, while a $\tau$ of 98 can only result from 98--0, 99--1 and
  100--1.}  Therefore, low values of $\tau$ represent more data for a
given $T$, while larger values represent fewer data points. If
particle motion is purely diffusive with zero mean then, provided that
the distance moved by particles in time $\tau$ is smaller than the
distance over which the diffusion coefficient varies, we would have
\begin{equation}
  D(r,\tau)=\frac{Q (r,\tau)}{2n(r,\tau)\tau} 
\end{equation}
where $D(r,\tau)$ is the diffusion coefficient at $r$.  If there is a
mean velocity field in addition to diffusion then we would have
\begin{equation}
  D(r,\tau) = \frac{Q(r,\tau)}{2\tau n(r,\tau)} - \frac{P(r,\tau)^2}{2\tau n(r,\tau)^2}.
\end{equation}
If the motion is purely diffusive then $D$ will not depend on $\tau$,
but if there is a more complicated background velocity field the
situation is more complicated as we see in the next section.

\subsection{Wave Motion with Diffusion}
 In order to under understand the motion of a particle in a
  wave acted on by diffusion we consider a particle moving in a wave
field with velocity given by $\omega A\cos(\phi+\omega t)$ and with
random fluctuations given by $N(t)$ so that its equation of motion is
\begin{equation}
  \frac{dx}{dt} = \omega A\cos(\phi+\omega t) + N(t).
\end{equation} 
This can be integated to give
\begin{equation}
  x(t) =  A\sin(\phi+\omega t) + \int_0^t N(s)\,ds.
\end{equation} 
Now we assume that $N(t)$ is Gaussian white noise with standard
deviation $\sigma$ so that $\int_0^t N(s)ds=\sigma W(t)$, where $W(t)$
is a Wiener process \citep{weiner}. We consider an ensemble of
trajectories and average uniformly over the phases $\phi$ and the
Wiener process using $\langle W(t) \rangle=0$ and
$\langle W(t)W(s) \rangle=\min(s,t)$.  Then we have, the
  expectation (denoted with $\langle \cdots \rangle$) of the position
\begin{equation}
  \langle x(t)\rangle =  0
\end{equation} 
and
\begin{equation}
  \langle [x(t+\tau)-x(t)]^2\rangle = A^2[1-\cos(\omega \tau)] + \sigma^2 \tau.
\end{equation} 
This function of $\tau$ is the sum of a linear function
  ($A^2+\sigma^2 \tau$) and an an oscillatory function
  $-A^2\cos(\omega \tau)$). At large times ($\tau\omega \gg 1$) there
  is a linear trend with gradient $\sigma^2$. From this gradient we can then extract
the effective diffusion coefficient $D=\sigma^2/2$.
\subsection{Application to Numerical Simulations}

We carry out this procedure in post-processing.  That is, given saved
velocity data from our hydrodynamic simulations at time intervals $t$,
we introduce $N$ particles into our numerical simulation and measure
$D$ using the above procedure.  Since the procedure is done in
post-processing, we can include an arbitrary number of particles to
check numerical convergence, we can also vary the type of particle
interpolation and the number of timesteps over which we track the
particles.  Fig.~\ref{snaps} shows time snapshots of the vorticity and
particle positions (only 15000 shown) at four different times.

We checked our procedure against a hydrodynamic simulation with
particles run within the simulation in order to confirm our velocity
data was taken at fine enough time resolution.  We find that in order
to measure a diffusion coefficient within the radiation zone, we need
long timeline data, but not very finely spaced.  However, to smoothly
resolve the convective-radiative interface we need finely spaced time
data over shorter timeline.  In this letter we are concerned with the
diffusion profile within the radiation zone, so we integrate over long
timelines with longer time steps and note that our convective and
overshoot profile is not well resolved nor well described as a
diffusive process.
\section{Particle Trajectories and  Diffusion}

\begin{figure}
\centering
\includegraphics[width=0.95\columnwidth]{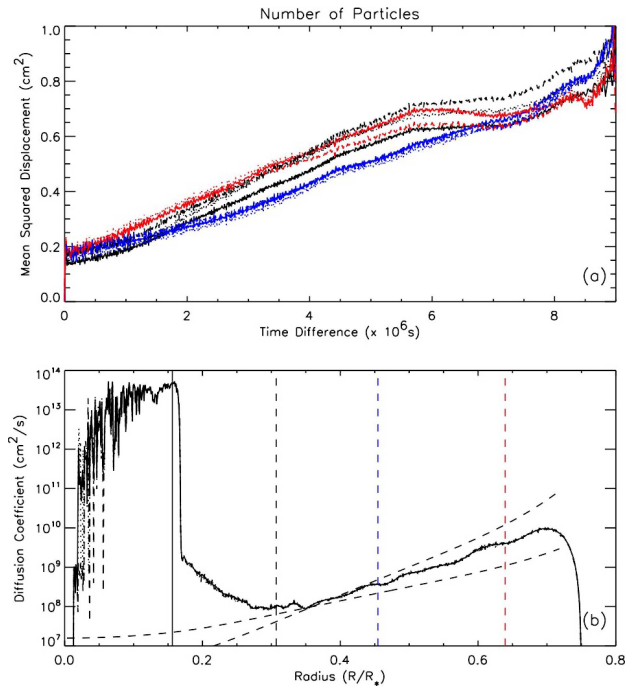}
\caption{Particle Diffusion.  (a) Mean squared displacement as a
  function of time difference $\tau$, with total time tracked,
    T=10$^7$s.  Different colors indicate different radii (as
  indicated by the vertical dashed lines of the same color in (b)).
  (a) Different line types represent different numbers of particles
  with 1 million (solid line), 100,000 (dashed line) and 50,000
  paricles (dotted line).  (b) Shows the diffusion coefficient as a
  function of radius with different line types as in (a).  There is no
  perceptible difference in diffusion coefficient within the radiative
  zone for varying particle number, indicating convergence. Dashed
  black lines represent $\rho^{-1/2}$ and $\rho^{-1}$.}
\label{msq2}
\end{figure}

The first question to address is whether the movement of particles due
to the action of waves can be treated as a diffusive process. To test
this we plot the mean squared displacement as a function of $\tau$ in
Fig.~\ref{msq2} (a) with different line types of the same color
showing dependencies on the number of particles used and different
colors showing different radii (see caption for details). There we see
that at low $\tau$ the particles indeed behave like diffusion (the
mean squared displacement is a linear function of $\tau$).  However at
larger $\tau$, there is not enough data to confirm the diffusive
nature.  We also see that different radial levels have different
slopes.
 
Using low values of $\tau$ (2.5 $\times 10^6$s, for which we have
sufficient data), we compute the diffusion coefficient as a function
of radius, which is shown in Fig.~\ref{msq2}(b), again with different
line types showing different numbers of particles.  The vertical black
solid line shows the convective-radiative interface, while the
vertical black, blue and red dashed lines show the radial positions
shown in (a). There we see that the overall radial profile of
diffusion coefficient within the radiation zone is robust to
variations in particle number, thus demonstrating convergence.  We
also see that there is a rapid transition from the behavior in the
convection zone to that in the radiative region.  We note that, while
there is a diffusion coefficient (plotted and measured) within the
convection zone and overshoot region, the behavior in those regions is
generally \textit{not} diffusive, particularly in the convection zone.
The mean squared displacement as a function of $\tau$ does not lie on
a straight line.  Therefore, applying a diffusion coefficient in these
regions is not appropriate.  However, we can see from Fig.~\ref{snaps}
that particles mix much more rapidly within the convection zone than
in the radiative region, as expected.

In general, within the radiation zone, the amplitude of the diffusion
coefficient rises with increasing distance from the convection zone.
There is decay just outside the convection zone due to the fact that
our thermal diffusivity is constant, rather than a function of radius
(we discuss this in Section 4). The radial increase is proportional to
a factor between $\rho^{-1}$ and $\rho^{-1/2}$ which we show in black
dashed lines and which we also discuss in Section 4. Therefore, we
conclude that: 1) IGW mixing in the radiation zone can be treated as a
diffusive process and 2) the radial profile of diffusion is robust and
is proportional to a function between $\rho^{-1}$ and $\rho^{-1/2}$.

\section{Parametrization of the Diffusion profile}

\begin{figure}
  \centering
  \includegraphics[width=0.95\columnwidth]{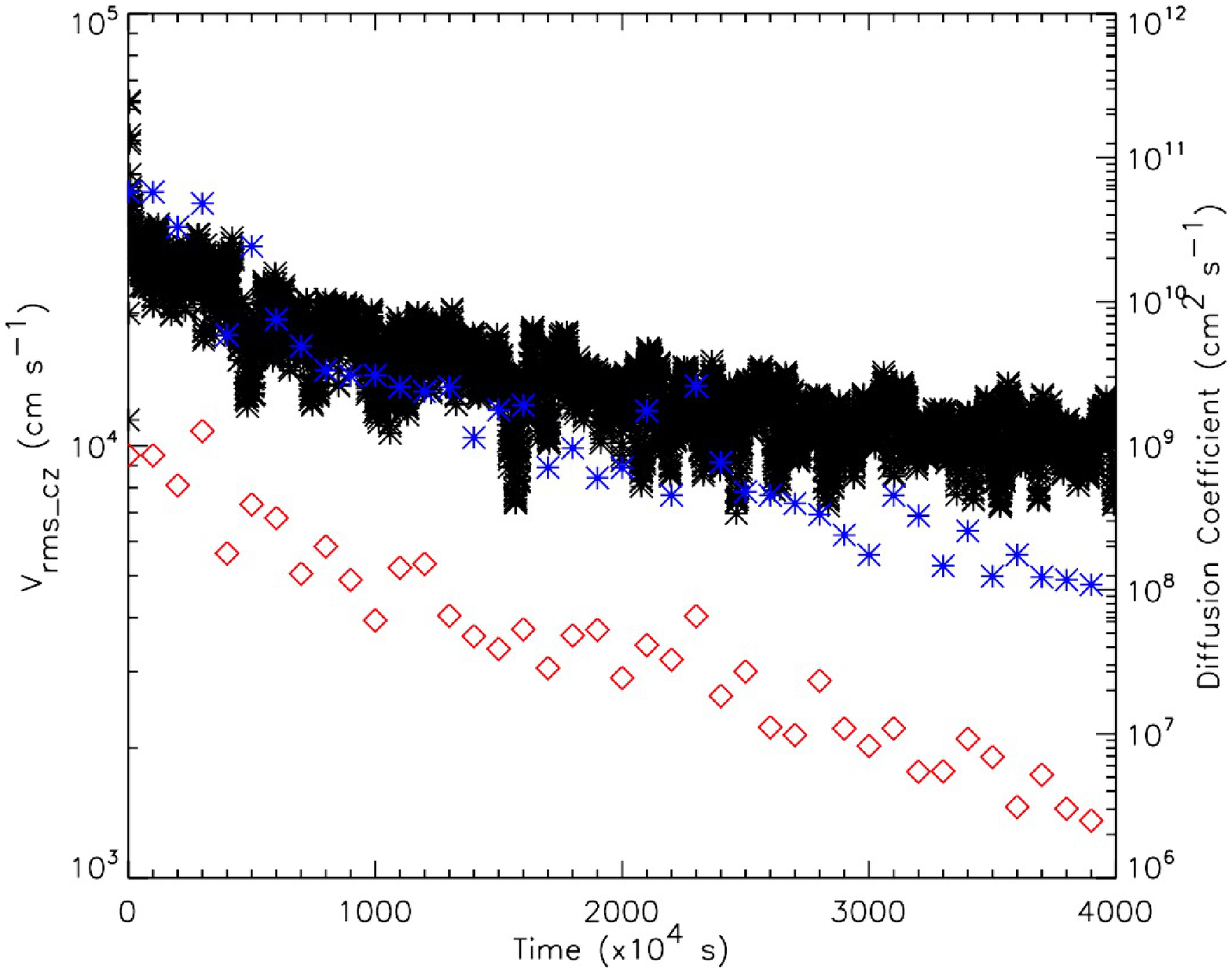}
  \caption{Root mean squared velocity (black asterisks) and diffusion
    coefficients versus time.  Diffusion coefficients are at 0.375
    R$_{\ast}$ (red diamonds) and at 0.675 R$_{\ast}$ (blue asterisks)
    and are calculated using Equation (6) and a particle
    tracking time, T=10$^{6}$s.} 
  \label{velvsdiff}
\end{figure}

In order to find a useful parametrization of IGW to be used in
one-dimensional (1D) stellar evolution models, we would like to know
both the amplitude and the radial profile of diffusion from our
simulations.  We find that the amplitude of the diffusion coefficient
we obtain within the radiation zone is correlated with the root mean
squared (rms) velocity \textit{within the convective zone}. In the
model presented, the root mean squared velocity in the convection zone
is decreasing in time.  This is an artifact of the fact that we force
the convection through a superadiabaticity, which is reduced due to
efficient convection.  In previous simulations we have a used a
forcing term to drive convection in order to avoid this, but here we
allowed this to investigate the role of convective velocities. We see
how this decay affects the diffusion coefficient in the radiation zone
in Fig.~\ref{velvsdiff} --- as convective velocities decrease, the
diffusion coefficient decreases. As we show later, this is because the
rms velocity within the convection zone sets the wave amplitudes
within the radiation zone.  Within a 1D stellar evolution code the rms
velocity could be approximated as the convective velocity given by
mixing length theory (MLT).

\begin{figure}
\centering
\includegraphics[width=0.95\columnwidth]{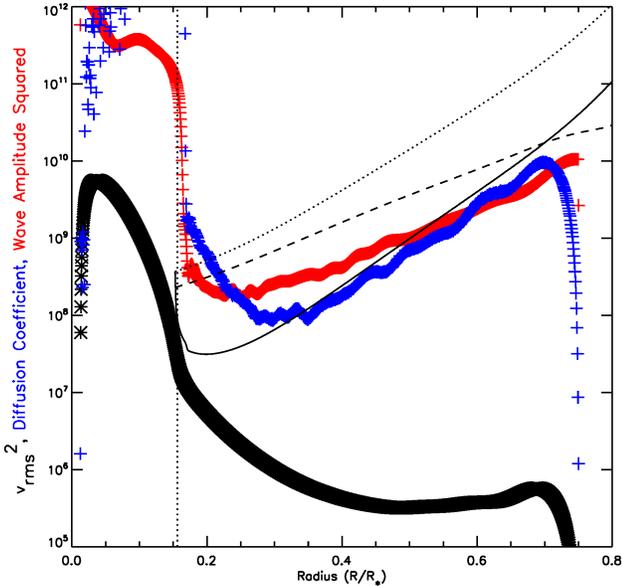}
\caption{Root mean squared velocity (black asterisks, cm$^2$s$^{-2}$),
  wave amplitude squared (red crosses,cm$^2$s$^{-2}$) and diffusion
  coefficient (blue crosses,cm$^2$s$^{-1}$) versus radius.  Wave
  amplitude squared is calculated using the larges scale wave
  ($k_h$=1) and all frequencies.  Black lines represent the wave
  amplitude squared calculated using Eqn.~(12).  The solid line
  assumes $\kappa=2\times 10^{12}$ cm$^2$/s as in the numerical
  simulation and a frequency spectrum at generation of velocities
  $\propto \omega^{-1}$, while the dotted and dashed lines use
  $\kappa$ from MESA and a frequency spectrum at generation of
  velocities $\propto \omega^{-1}, \omega^{-3}$, respectively.}
\label{waveamp}
\end{figure}
Independent of the amplitude of the diffusion coefficient, we find
that the radial profile of diffusion seen in Fig.~\ref{msq2} is robust
across choices of numerical diffusion (the thermal and viscous
diffusivity are each varied by an order of magnitude) and convective
velocities.  Now we would like to determine what sets the radial
profile of diffusion. In Fig~\ref{waveamp} we plot the rms velocity
(black asterisks), the wave amplitude squared (red crosses) and the
diffusion coefficient (blue crosses) as a function of radius.  The
wave amplitude is calculated with Eqn.~(12) using the largest scale
wave and all frequencies we calculate (up to 500\,$\mu$Hz).  In
general, frequencies above $\sim$40\,$\mu$Hz do not contribute as
their generation amplitudes are too low.  We see that the diffusion
profile closely correlates with the wave amplitude squared. This
numerical finding is consistent with the theoretical prediction that
diffusion is the autocorrelation of the lagrangian velocity field. In
this case the velocity field is the wave amplitude, which is
correlated over long timescales.  The diffusion profile is
\textit{not} correlated with the local rms velocity, which is again
consistent with with the theoretical expectation that diffusion is the
autocorrelation of the velocity field, as the local rms velocities are
not correlated over long timescales and hence have negligible
autocorrelation.
  
Due to numerical constraints (dimensionality, diffusion coefficients)
it is likely the IGW amplitudes in our simulations are not
realistic. Using our numerical result that diffusion is proportional
to the wave amplitude squared, we now turn to theoretical models for a
parametrization of wave diffusion in stellar interiors. The amplitude
of an internal gravity wave (IGW) depends on the wave driving at the
convective-radiative interface, the radiative damping of the wave and
the density stratification.  The wave driving is directly correlated
with the rms velocity within the convection zone, $v_{rms-cz}$
\citep{gls91}.  Within the radiation zone the wave is damped by
thermal diffusion \citep{ktz99} and amplified by the density
stratification (due to conservation of pseudomomentum
\citep{buhler09}). Therefore, the wave amplitude is determined from
the simple function
\begin{equation}
v_{wave}=v_{rms-cz} \,\left(\frac{\rho\left(r\right)}{\rho_{tcz}}\right)^{-1/2} e^{-\tau\left(\omega,k_h,r\right)}
\end{equation}
where $\rho_{tcz}$ is the density at the top of the convection zone.
From Fig.~\ref{waveamp} we estimate the diffusion due to IGW as
\begin{equation}
D_{mix}=\mathcal{A} v_{wave}^2\left(\omega,k_h,r\right)
\end{equation}
where $\mathcal{A}$ has units of s and $\tau\left(\omega,k_h,r\right)$ is the damping ``optical depth''
of a wave defined in \cite{ktz99} as:
\begin{equation}
\tau\left(\omega,k_h,r\right)=\int_{r_{tcz}}^{r} \frac{\kappa k_h^3
  N^3}{\omega^4 \left(2\pi r\right)^3} \,dr
\end{equation}
where $\kappa$ is the radiative diffusivity, $k_h$ is the horizontal
wavenumber of the wave, $r_{tcz}$ is the radius at the top of the
convection zone, $N$ is the Brunt-V\"ais\"al\"a frequency and $\omega$
is the frequency of the wave. For simplicity, we have assumed no
doppler shift in the frequencies of the waves.  $\mathcal{A}$ is an
unknown constant, which is $\sim$1s in our models.  Although the
precise value is unknown, we don't expect it to vary significantly
from this (see Discussion).

The initial decay of diffusion outside the convection zone is due to
the fact that we use a constant thermal diffusion coefficient,
$\kappa$, rather than the stellar radiative value. The numerical value
used is $2\times 10^{12}$ cm$^2$/s throughout, while the value in the
star varies from $10^7$ cm$^2$/s at the convective-radiative interface
to $10^{15}$ cm$^2$/s at the surface of the star.  Therefore,
throughout our computed radiative zone we are damping the waves more
than they would be damped in the stellar interior, this is
particularly true just outside the convection zone and is the reason
for the initial decay seen in the wave amplitude squared (red line,
Fig.~\ref{waveamp}).  This is demonstrated by the black lines which
show the wave amplitude computed using Eqn.~(12) with the value of
$\kappa$ used in our simulation (solid line) and the values of
$\kappa$ from the stellar model (dashed and dotted lines).  For this
simple calculation we have assumed that the waves are linear,
non-interacting and that the wave amplitude at the
convective-radiative interface is half the rms velocity.\footnote{This
  amplitude comes from a simple calculation assuming $F_w\sim M F_c$
  \citep{lecoanet13}, where M is the Mach number of the convection and
  $F_c$ and $F_w$ are the convective and wave fluxes, respectively.
  We assume that $F_c, F_w \sim v_c^3, v_w^3$, therefore,
  $v_w\sim M^{1/3}v_c$.  Assuming $M\sim 0.1$ then
  $v_w\sim 0.47 v_c$.}  Predictions for the frequency spectra of waves
generated by convection at a given wavenumber range from $\omega^{-3}$
\citep{ktz99,lecoanet13} to $\omega^{-1}$ \citep{rg10,rg13}.
Therefore, the dashed and dotted lines represent a frequency
generation spectra of velocity proportional to $\omega^{-3}$ and
$\omega^{-1}$ respectively, to cover that range. We integrate over the
same frequency range and scales as for the numerical results (red
asterisks in Fig.~\ref{waveamp}).  We see that the initial decay
outside the convection zone is purely an artifact of enhanced
diffusion for numerical purposes and is not physical.

\section{Discussion}

In this letter we have demonstrated that the mixing by IGW within
radiative regions can be treated as a diffusive process.  We have
further shown that the local amplitude of the diffusion coefficient
depends on the local wave amplitude.  The wave amplitude, in turn,
depends on the convective forcing ($v_{rms-cz}$), the thermal damping
of the wave and the density stratification in a simple way.
Therefore, a prescription for the diffusion coefficient due to mixing
by IGW can be easily implemented using Eqns.~(12--14) assuming MLT
velocities for the rms velocity and all other parameters determined by
the stellar model.  The one parameter that is left is $\mathcal{A}$.
While this value is $\sim$1 in our simulations, its precise value may
depend on the stellar viscosity/thermal diffusivity, rotation and the
dimensionality.  Thermal diffusivity is already accounted for in (13).
Since viscosity is enhanced in our simulations one would expect
$\mathcal{A}=1$ to be a lower limit.  However, one does not expect
viscosity to play a role in the propagation of linear waves, so the
prescription in (13) with $\mathcal{A}=1$ likely still holds. In the
case of rotation, fast rotation would likely reduce the wave amplitude
and hence, $\mathcal{A}=1$ would be considered an upper limit, but
that is likely a small effect.  It is unclear what effect our reduced
dimensionality has on the waves.  Assuming that simple one dimensional
wave propagation is sufficient, the dimensionality likely only affects
the wave generation spectrum.  Taking all this into account, the best
approach would be to assume $\mathcal{A}=1$ and vary the wave
generation spectrum incorporated through Eqn.~(14).  Then the one
parameter of the model would be the exponent of the frequency spectrum
of wave generation.

Given numerical limitations, for the forseeable future the most
reliable constraints on $\mathcal{A}$ would come from comparisons
between theoretical models using this prescription and observations of
slowly rotating intermediate mass stars.  Asteroseismic inversions
could place constraints on near-core mixing \citep{ehsan15,ehsan16},
while spectroscopic observations may place constraints on subsurface
mixing. Simultaneous comparisons between theoretical evolution models,
spectroscopic and asteroseismic data could provide constraints on the
entire diffusion profile and indeed may help place constraints on the
wave generation spectrum.

Finally, our simulations only extend to 0.7R$_{\ast}$.  Extending our
linear calculations using Eqn.~(12) shows that the wave amplitude
continues to increase until just beneath the stellar surface. It is
likely that these waves break \citep{rg13} and hence the surface
diffusion coefficient may be enhanced (due to turbulent mixing) beyond
what is expected from linear wave behavior.  Numerical simulations
attempting to resolve the wave dynamics near the stellar surface will
be forthcoming.

\bibliographystyle{apj}


\begin{thebibliography}{35}
\expandafter\ifx\csname natexlab\endcsname\relax\def\natexlab#1{#1}\fi

\bibitem[{Aerts {et~al.}(2014)Aerts, Molenberghs, Kenward, \& Neiner}]{aerts14}
Aerts, C., Molenberghs, G., Kenward, M., \& Neiner, C. 2014, The Astrophysical
  Journal, 781, 88

\bibitem[{Alvan {et~al.}(2015)Alvan, Strugarek, Brun, Mathis, \&
  Garcia}]{alvan15}
Alvan, L., Strugarek, A., Brun, A., Mathis, S., \& Garcia, R. 2015, Astronomy
  and Astrophysics, 581, 112

\bibitem[{Augustson {et~al.}(2016)Augustson, Brun, \& Toomre}]{augustson16}
Augustson, K., Brun, A., \& Toomre, J. 2016, ApJ, 829, 92

\bibitem[{Bohm-Vitense(1958)}]{bv58}
Bohm-Vitense, E. 1958, Z. Astrophysics, 46, 108

\bibitem[{Brott {et~al.}(2011)Brott, Evans, de~Koter, Langer, Dufton,
  Cantiello, \& Trundle}]{brott11}
Brott, I., Evans, C., de~Koter, A., Langer, N., Dufton, P., Cantiello, M., \&
  Trundle, C. 2011, Astronomy and Astrophysics, 530, 116

\bibitem[{Browning {et~al.}(2004)Browning, Brun, \& Toomre}]{browning04}
Browning, M., Brun, A., \& Toomre, J. 2004, ApJ, 601, 512

\bibitem[{Buhler(2009)}]{buhler09}
Buhler, O. 2009 (Cambridge Monographs on Mechanics, Cambridge University Press)

\bibitem[{Charbonnel \& Talon(2005)}]{charbonnel05}
Charbonnel, C., \& Talon, S. 2005, Science, 309, 2189

\bibitem[{Doob(1953)}]{weiner}
Doob, J. 1953, 101

\bibitem[{Eddington(1925)}]{eddington25}
Eddington, A. 1925, The Observatory, 48, 73

\bibitem[{Edelmann {et~al.}(2017)Edelmann, Roepke, R.Hirschi, Cyril, \&
  S.Jones}]{edelmann17}
Edelmann, P., Roepke, F., R.Hirschi, Cyril, G., \& S.Jones. 2017, Astronomy and
  Astrophysics, 224, 245

\bibitem[{Eggenberger {et~al.}(2017)Eggenberger, Lagarde, Miglio, Montalban,
  Ekstrom, Georgy, Meynet, Salmon, Ceillier, Garcia, Mathis, Deheuvels, Maeder,
  den Hartogh, \& Hirschi}]{eggenberger17}
Eggenberger, P., {et~al.} 2017, Astronomy and Astrophysics, 599, 18

\bibitem[{Garcia-Lopez \& Spruit(1991)}]{gls91}
Garcia-Lopez, R., \& Spruit, H. 1991, ApJ, 377, 268

\bibitem[{Gough(1969)}]{go69}
Gough, D. 1969, Journal of Atmospheric Sciences, 26, 448

\bibitem[{Heger {et~al.}(2000)Heger, Langer, \& Woosley}]{heger00}
Heger, A., Langer, N., \& Woosley, S. 2000, The Astrophysical Journal, 528, 368

\bibitem[{Kippenhahn {et~al.}(2012)Kippenhahn, Weigert, \& Weiss}]{kippenhahn}
Kippenhahn, R., Weigert, A., \& Weiss, A. 2012

\bibitem[{Kumar {et~al.}(1999)Kumar, Talon, \& Zahn}]{ktz99}
Kumar, P., Talon, S., \& Zahn, J. 1999, ApJ, 520, 859

\bibitem[{Lecoanet \& Quataert(2013)}]{lecoanet13}
Lecoanet, D., \& Quataert, E. 2013, Monthly Notices of the Royal Astronomical
  Socity, 430, 2363

\bibitem[{Moravveji {et~al.}(2015)Moravveji, Aerts, Papics, Triana, \&
  Vandoren}]{ehsan15}
Moravveji, E., Aerts, C., Papics, P., Triana, S., \& Vandoren, B. 2015,
  Astronomy and Astrophysics, 580, 27

\bibitem[{Moravveji {et~al.}(2016)Moravveji, Townsend, Aerts, \&
  Mathis}]{ehsan16}
Moravveji, E., Townsend, R., Aerts, C., \& Mathis, S. 2016, The Astrophysical
  Journal, 823, 130

\bibitem[{Paxton {et~al.}(2010)Paxton, Bildstein, Dotter, Herwig, Lesaffre, \&
  Timmes}]{paxton10}
Paxton, B., Bildstein, L., Dotter, A., Herwig, F., Lesaffre, P., \& Timmes, F.
  2010, The Astrophysical Journal Supplement Series, 192, 3

\bibitem[{Paxton {et~al.}(2015)Paxton, Marchant, Schwab, Bauer, Bildsten,
  Cantiello, Dessart, Farmer, Hu, \& Langer}]{paxton15}
Paxton, B., {et~al.} 2015, The Astrophysical Journal Supplement Series, 220, 15

\bibitem[{Pinsonneault(1997)}]{pin97}
Pinsonneault, M. 1997, Annual Review of Astronomy and Astrophysics, 35, 557

\bibitem[{Press \& Rybicki(1981)}]{pressryb81}
Press, W., \& Rybicki, G. 1981, ApJ, 248, 751

\bibitem[{Renzini(1987)}]{renzini87}
Renzini, A. 1987, Astronomy and Astrophysics, 188, 49

\bibitem[{Rogers(2015)}]{rg15}
Rogers, T. 2015, ApJL, 815, L30

\bibitem[{Rogers \& Glatzmaier(2005)}]{rg05b}
Rogers, T., \& Glatzmaier, G. 2005, MNRAS, 364, 1135

\bibitem[{Rogers {et~al.}(2012)Rogers, Lin, \& Lau}]{rg12}
Rogers, T., Lin, D., \& Lau, H. 2012, ApJL, 758, 6

\bibitem[{Rogers {et~al.}(2013)Rogers, Lin, McElwaine, \& Lau}]{rg13}
Rogers, T., Lin, D., McElwaine, J., \& Lau, H. 2013, ApJ, 772, 21

\bibitem[{Rogers \& MacGregor(2010)}]{rg10}
Rogers, T., \& MacGregor, K. 2010, Monthly Notices of the Royal Astronomical
  Society, 410, 946

\bibitem[{Spruit(2002)}]{spruit02}
Spruit, H. 2002, Astronomy and Astrophysics, 381, 923

\bibitem[{Sweet(1950)}]{sweet50}
Sweet, P. 1950, Monthly Notices of the Royal Astronomical Society, 110, 548

\bibitem[{Talon \& Charbonnel(2005)}]{talon05}
Talon, S., \& Charbonnel, C. 2005, Astronomy and Astrophysics, 440, 981

\bibitem[{Vogt(1925)}]{Vogt25}
Vogt, H. 1925, Astronomische Nachrichten, 224, 245

\bibitem[{Zahn(1991)}]{zahn91}
Zahn, J. 1991, Astronomy and Astrophysics, 252, 179


\end{thebibliography}

\acknowledgments 
Support for this research was provided by an STFC Grant ST/L005549/1 and NASA
grant NNX17AB92G.  Computing was carried out on Pleiades at NASA Ames.
T. Rogers thanks Conny Aerts, May Gade-Pederson and Timothy Van Reeth for
useful conversations leading to the development of this manuscript.
We both thank an anonymous referee for very useful suggestions which
greatly improved the manuscript.

\end{document}